\documentclass[10pt,letterpaper,conference]{IEEEtran}
\usepackage{units}
\usepackage{textcomp}
\usepackage{amsmath}
\usepackage{amssymb}
\usepackage{graphicx}

\makeatletter

\usepackage{cite}

\makeatother

\begin{document}

\title{Coherent 100G Nonlinear Compensation with Single-Step Digital Backpropagation}

\author{\IEEEauthorblockN{M. Secondini, S. Rommel, F. Fresi, E. Forestieri}
\IEEEauthorblockA{TeCIP Institute, Scuola Superiore Sant'Anna\\
I-56124 Pisa, Italy.}
\and
\IEEEauthorblockN{G. Meloni, L. Pot\`{\i}}
\IEEEauthorblockA{National Laboratory of Photonics Networks, CNIT\\
 I-56124 Pisa, Italy.}}
\maketitle
\begin{abstract}
Enhanced-SSFM digital backpropagation (DBP) is experimentally demonstrated
and compared to conventional DBP. A 112\,Gb/s PM-QPSK signal is transmitted
over a 3200\,km dispersion-unmanaged link. The intradyne coherent
receiver includes single-step digital backpropagation based on the
enhanced-SSFM algorithm. In comparison, conventional DBP requires
twenty steps to achieve the same performance. An analysis of the computational
complexity and structure of the two algorithms reveals that the overall
complexity and power consumption of DBP are reduced by a factor of
16 with respect to a conventional implementation, while the computation
time is reduced by a factor of 20. As a result, the proposed algorithm
enables a practical and effective implementation of DBP in real-time
optical receivers, with only a moderate increase of the computational
complexity, power consumption, and latency with respect to a simple
feed-forward equalizer for dispersion compensation. \end{abstract}

\section{Introduction}

Digital backpropagation (DBP) is one of the most studied strategies
to counteract nonlinearities by channel inversion \cite{Ip:JLT08}.
Due to both technological and practical reasons, DBP is typically
used only to compensate for intra-channel nonlinearity. Though most
effective for combating nonlinear signal-signal interactions, DBP
is also a key element for the realization of nearly optimum detectors
(accounting also for signal-noise interaction) when combined to a
Viterbi processor for maximum likelihood sequence detection \cite{Marsella:JLT14},
or applied to particle filtering for stochastic backpropagation \cite{Naga14}.
In practice, DBP is implemented through the split-step Fourier method
(SSFM), probably the most efficient numerical method known to simulate
fiber-optic propagation \cite{Taha84}. The SSFM allows for complexity
versus accuracy trade-off by adjusting the number of steps $N_{s}$
and takes advantage of the high computational efficiency of the fast
Fourier transform algorithm. Nevertheless, the computational complexity,
latency, and power consumption required by DBP in typical long-haul
systems are significantly higher than those required by other digital
signal processing blocks (e.g., linear equalizers) and still pose
some difficulties to its implementation in a real-time digital receiver.
Several approaches have been proposed to obtain good trade-offs between
complexity and performance \cite{Goldfarb:OE2009,Asif:OE2010}. Based
on heuristic approaches, some modifications of the SSFM algorithm
have been also proposed to reduce complexity without sacrificing accuracy
\cite{Du:OE2010,Li:OFC11,Ip:OFC11,Rafique:OE2011}. However, some
approximate solutions of the nonlinear Schr\"{o}dinger equation are
available in the literature and could be exploited for improving the
SSFM algorithm \cite{Peddanarappagari97,Van:JLT0702,Ciar:PTL0105,For:TIWDC04}.
While the Volterra series \cite{Peddanarappagari97} and regular perturbation
\cite{Van:JLT0702} approaches describe the nonlinearity as an additive
perturbation and do not appear to be suitable, the logarithmic approximation
\cite{Ciar:PTL0105,For:TIWDC04} gives an expression which is closer
to that used in the SSFM method for approximating the nonlinear propagation
in a piece of fiber but more accurate. Based on the logarithmic perturbation
technique, an enhanced SSFM (ESSFM) has been recently proposed and,
through numerical simulations, it was shown to have a one order of
magnitude lower complexity compared to the standard SSFM for a prescribed
accuracy \cite{Secondini:ECOC14}. 

In this work, we extend the ESSFM algorithm to account for the propagation
of a polarization-multiplexed signal and experimentally demonstrate
its effectiveness for the implementation of DBP within a coherent
optical receiver. In particular, we compare the SSFM and ESSFM algorithms
to backpropagate a 112\,Gb/s PM-QPSK signal through a 3200\,km dispersion-unmanaged
link, showing that the ESSFM provides a significant reduction of complexity,
latency, and power consumption. The paper is organized as follows.
In Section~II, we describe the ESSFM algorithm. In Section~III,
we investigate the computational complexity, computational time (latency),
and power consumption of the proposed ESSFM algorithm and compare
them to those of the conventional SSFM and of a simple feed-forward
equalizer (FFE) for dispersion compensation. In Section IV, we show
the experimental results and the actual improvements obtained by employing
the ESSFM. Finally, in Section V, we draw the conclusions.

\section{Enhanced split-step Fourier method}

The propagation of a single-polarization optical signal through a
fiber-optic link in the presence of chromatic dispersion, Kerr nonlinearity,
and attenuation is governed by the nonlinear Schr\"{o}dinger equation,
which can be numerically solved by means of the SSFM algorithm. According
to the SSFM, the link is divided into $N_{s}$ small segments (steps).
Each step is further divided into two sub-steps: a linear sub-step,
accounting for chromatic dispersion, and a nonlinear sub-step, accounting
for a nonlinear phase rotation proportional to the signal intensity
(Kerr nonlinearity). When considering polarization-multiplexed signals,
the nonlinear Schr\"{o}dinger equation is replaced by the Manakov
equation \cite{Men:JLT06}. In this case, the SSFM can be still employed
by modifying the nonlinear sub-step to account for a nonlinear phase
rotation on each polarization that is proportional to the overall
signal intensity on both polarizations. In both cases, as processing
for the linear and nonlinear sub-steps takes place in frequency and
time domain, respectively, direct and inverse FFTs are used at each
step to switch between the time and frequency representation of the
signal. In practice, the propagation of a block of $N$ vector samples
$\{\mathbf{x}_{k}\}_{k=1}^{N}$ (where each vector $\mathbf{x}_{k}=(x_{k,1},x_{k,2})^{T}$
collects the $k$-th samples of the two signal polarizations) through
a generic step of length $\Delta z$, with dispersion coefficient
$\beta_{2}$, nonlinear coefficient $\gamma$, and attenuation coefficient
$\alpha$, entails performing the following four operations: \emph{(i)}
computation of the frequency components $\{\mathbf{X}_{k}\}_{k=1}^{N}$
through a pair of FFTs (one per each polarization); \emph{(ii)} computation
of the linear sub-step
\begin{equation}
\mathbf{Y}_{k}=\mathbf{X}_{k}e^{-j2\pi^{2}\beta_{2}f_{k}^{2}\Delta z},\quad k=1,\ldots,N\label{eq:passo_lineare}
\end{equation}
where $f_{k}$ is the frequency of the $k$-th component; \emph{(iii)}
computation of the time components $\{\mathbf{y}_{k}\}_{k=1}^{N}$
through a pair of inverse FFTs; \emph{(iv)} computation of the nonlinear
sub-step 
\begin{equation}
\mathbf{z}_{k}=\mathbf{y}_{k}e^{-j\gamma\Delta z_{\mathrm{eff}}|\mathbf{y}_{k}|^{2}},\quad k=1,\ldots,N\label{eq:passo_nonlineare_SSFM}
\end{equation}
$\Delta z_{\mathrm{eff}}=(1-e^{-\alpha\Delta z})/\alpha$ being the
effective length. The output sequence $\{\mathbf{z}_{k}\}_{k=1}^{N}$
becomes, in turn, the input to the next fiber segment and so on, until
the end of the link is reached.

The overall complexity of the SSFM algorithm is mainly driven by the
required $4N_{s}$ FFTs and can be reduced by employing the ESSFM
algorithm, which achieves the same accuracy as the SSFM with a lower
number of steps $N_{s}$ \cite{Secondini:ECOC14}. The main idea behind
the ESSFM is that of keeping the SSFM approach but modifying the nonlinear
sub-step (\ref{eq:passo_nonlineare_SSFM}) to account also for the
interaction between dispersion and nonlinearity along $\Delta z$.
In this way, $\Delta z$ can be increased (and, consequently, $N_{s}$
decreased) without affecting the overall accuracy. Of course, the
overall complexity is reduced only if the new term is less costly
than the spared FFTs. A more accurate expression for the nonlinear
sub-step is provided by the frequency-resolved logarithmic perturbation
(FRLP) method \cite{Sec:PTL2012}. In particular, it can be shown
that in the nonlinear step the signal undergoes a nonlinear phase
rotation that depends on a quadratic form of the signal samples \cite{Secondini:JLT13}.
By truncating the channel memory to the first $N_{c}$ past and future
samples, retaining only the diagonal terms of the quadratic form,
and averaging the FRLP coefficients over the signal bandwidth, results
in the modified nonlinear sub-step proposed in \cite{Secondini:ECOC14}
(for a single-polarization signal) which, in analogy to the SSFM case,
is simply extended to a polarization-multiplexed signal by considering
a nonlinear phase rotation on each polarization that is the sum of
the phase rotations induced by each polarization. The enhanced nonlinear
sub-step is thus expressed as 
\begin{equation}
\mathbf{z}_{k}=\mathbf{y}_{k}e^{-j\gamma\Delta z_{\mathrm{eff}}\left(c_{0}|\mathbf{y}_{k}|^{2}+\sum_{i=1}^{N_{c}}c_{i}(|\mathbf{y}_{k-i}|^{2}+|\mathbf{y}_{k+i}|^{2})\right)}\label{eq:passo_nonlineare_ESSFM}
\end{equation}
where $\{c_{i}\}_{i=0}^{N_{c}}$ are $N_{c}+1$ real coefficients.
Formally, (\ref{eq:passo_nonlineare_ESSFM}) is equal to the nonlinear
sub-step proposed in \cite{Secondini:ECOC14}, but replacing scalar
samples with vector samples. We also note that (\ref{eq:passo_nonlineare_ESSFM})
is similar to the nonlinear sub-step proposed in \cite{Li:OFC11,Rafique:OE2011}.
However, the coefficient values obtained through a logarithmic-perturbation
analysis or numerical optimization, as discussed in \cite{Secondini:ECOC14},
may be significantly different from the low-pass filter coefficients
employed in \cite{Li:OFC11,Rafique:OE2011}, thus providing a different
performance.

\section{Computational complexity, latency, and power consumption}

The hardest challenge for a real-time implementation of DBP is keeping
its complexity, latency, and power consumption within feasible values.
Though an accurate analysis of the computational complexity, latency,
and power consumption for a real-time implementation of the SSFM and
ESSFM algorithms is beyond the scope of this work---it depends on
the actual implementation of the FFT and of the exponential operation,
on the employed hardware, on the sampling rate, on the adopted precision,
and so on---here we want to show that the number of steps $N_{s}$
is a reasonable figure of merit to compare the two algorithms and
to provide a rough, yet meaningful, indication about their complexity,
latency, and power consumption.

When processing a long sequence of samples through the SSFM or ESSFM
algorithms, as required for instance when implementing DBP in a fiber-optic
transmission system, the overlap-and-save technique is typically employed
\cite{Stockham66,OpSch99}. The input sequence of samples is divided
into several overlapping blocks which are separately processed, and
the output sequence is then reconstructed by discarding the overlapping
samples. The number of overlapping samples should be at least equal
to the overall memory $M$ of the fiber-optic channel---which, for
dispersion-uncompensated links, can be approximated as $M\simeq2\pi|\beta_{2}|LB^{2}$,
where $\beta_{2}$ is the fiber dispersion parameter, $L$ the link
length, and $B$ the signal bandwidth (assumed equal to the sampling
rate)---while the block length $N$ should be optimized to minimize
the computational cost per propagated sample. The propagation of each
block of samples through each step of fiber requires: the computation
of four FFTs (a pair of direct and inverse FFTs per polarization)
of $N$ complex samples (about $8N\log_{2}N$ real multiplications
and $8N\log_{2}N$ real additions)%
\footnote{We consider the classical Cooley-Tukey radix-2 FFT algorithm \cite{CoolTuk65}
and assume that each complex multiplication requires 4 real multiplications
and 2 real additions. Though slightly more efficient implementations
are possible, this provides a reasonable indication of the required
operations. Moreover, we assume that all fixed quantities (e.g., $\gamma\Delta zc_{i}$
or $\exp(-j2\pi^{2}\beta_{2}f_{k}^{2}\Delta z)$) are precalculated
and that the complex exponential in (\ref{eq:passo_nonlineare_ESSFM})
is evaluated by using a lookup table.%
}; the computation of the linear sub-step (\ref{eq:passo_lineare})
($8N$ real multiplications and $4N$ real additions); the computation
of the nonlinear sub-step (\ref{eq:passo_nonlineare_ESSFM}), which
in turn requires the computation of $N$ squared moduli ($4N$ real
multiplications and $3N$ real additions), their linear combination
($NN_{c}+N$ real multiplications and $2NN_{c}$ real additions),
and the nonlinear phase shift rotation ($8N$ real multiplications
and $4N$ real additions, neglecting the cost of the complex exponential).
Overall, considering that $M$ samples out of $N$ are discarded by
the overlap-and-save algorithm, the ESSFM algorithm requires $N(8\log_{2}N+21+N_{c})/(N-M)$
real multiplications and $N(8\log_{2}N+11+2N_{c})/(N-M)$ real additions
per step per received sample. The complexity of the SSFM is exactly
the same, with $N_{c}=0$.

It is useful to make a comparison with the complexity of a linear
feed-forward equalizer (FFE) for bulk dispersion compensation. This
is typically implemented in frequency domain and is practically equivalent
to a single step of the SSFM, in which only the linear sub-step is
considered \cite{Kus:JLT0908}: two parallel direct FFTs (one per
polarization) of $N$ complex samples, the linear sub-step (\ref{eq:passo_lineare}),
and two parallel inverse FFTs. Overall, the FFE requires $N(8\log_{2}N+8)/(N-M)$
real multiplications and $N(8\log_{2}N+4)/(N-M)$ real additions per
received sample.

\begin{figure}
\begin{centering}
\includegraphics[width=1\columnwidth]{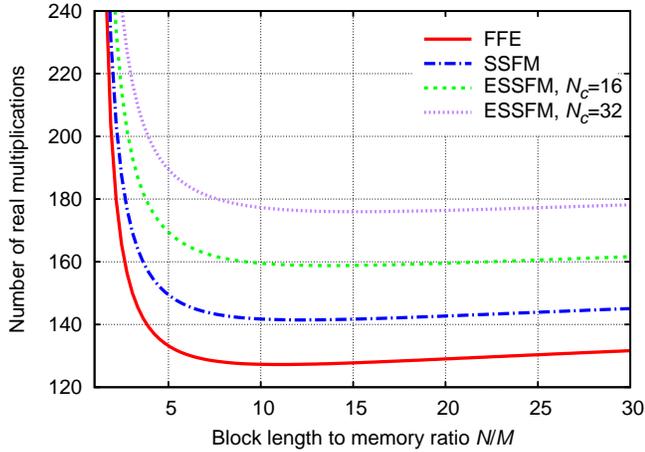}
\par\end{centering}

\caption{\label{fig1}Computational complexity per step as a function of the
block length $N$ for a channel memory of $M=1024$ samples and different
algorithms.}
\vspace*{1ex}
\end{figure}

 Given the memory $M$ of the channel, the block length $N$ can
be optimized to minimize the complexity (e.g., number of additions
and/or multiplications). As an example, considering a memory of $M=1024$
samples---due, for instance, to the propagation of a 50~GHz signal
through about 3000~km of standard single-mode fiber---Fig.\,\ref{fig1}
shows the number of real multiplications required by the FFE, by \emph{one
step} of the SSFM, and by \emph{one step} of the ESSFM (with $N_{c}=32$)
per each processed sample as a function of the ratio $N/M$. As it
is clear also from the expressions provided above, the optimum ratio
depends on the considered algorithm and on the value of $N_{c}$ (and
also on the value of $M$, assumed fixed in Fig.~\ref{fig1}). However,
it can be observed that by setting $N=8M$, one obtains nearly minimum
complexity in all the considered cases. Lower values of $N$ would
reduce latency, but at the expense of a significantly higher complexity.
On the other hand, higher values of $N$ would only slightly reduce
complexity, but at the expense of a higher latency. A similar result
is obtained also when considering the number of real additions and
different values of $M$ and $N_{c}$ (within a reasonable range of
practical interest). Therefore, in the following, we will always consider
$N=8M$.

\begin{figure}
\begin{centering}
\includegraphics[width=1\columnwidth]{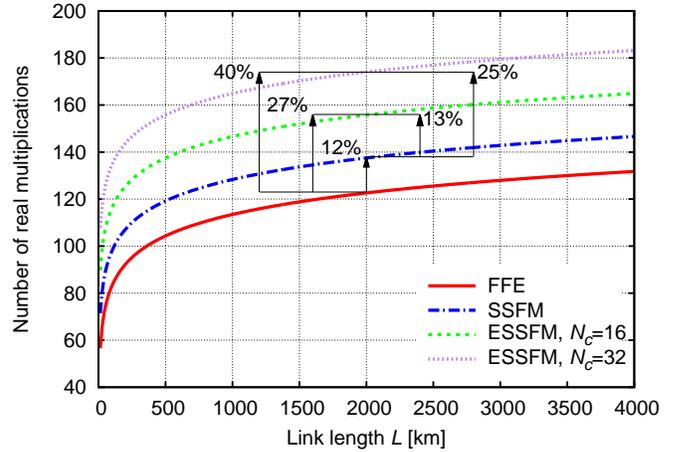}
\par\end{centering}

\caption{\label{fig2}Computational complexity per step as a function of the
link length for a block length $N=8M$ and different algorithms.}
\vspace*{1ex}
\end{figure}

Given this choice, it is interesting to compare the complexity of
the various algorithms and see how it changes with link length. Fig.~\ref{fig2}
reports the number of real multiplications per processed sample for
the FFE, SSFM, and ESSFM with different values of $N_{c}$ as a function
of the link length $L$. A standard single-mode fiber ($\beta_{2}=\unit[21]{ps^{2}/km}$)
and a 50~GHz signal bandwidth are considered. Note, again, that only
\emph{one step} of SSFM or ESSFM is considered, and that their overall
complexity is thus obtained by multiplying the values in Fig.\,\ref{fig1}
or \ref{fig2} by the total number of steps $N_{s}$. As shown in
the figure, at 2000~km each step of the ESSFM with $N_{c}=32$ is
about 25\% more complex than the SSFM and 40\% more complex than the
FFE. These figures remain almost unchanged at longer distances, slightly
increase at shorter distances, and decrease when considering a lower
number of coefficients $N_{c}$. 

When considering a real-time implementation of the algorithms in an
optical receiver, the resulting power consumption is an important
figure of merit. Power consumption depends on the actual hardware
implementation, processing rate, and considered technology and its
accurate and absolute estimate is beyond the scope of this work. However,
we can reasonably assume that, once all these parameters are fixed,
power consumption scales approximately as computational complexity.
Therefore, we can use Fig.~\ref{fig2} also to compare the power
consumption of the various algorithms and infer that \emph{each step}
of the ESSFM (with $N_{c}=32$) requires about 25\% more power than
\emph{each step} of the SSFM, and about 40\% more power than the FFE.

\begin{figure*}
\centering{}\includegraphics[width=1\textwidth]{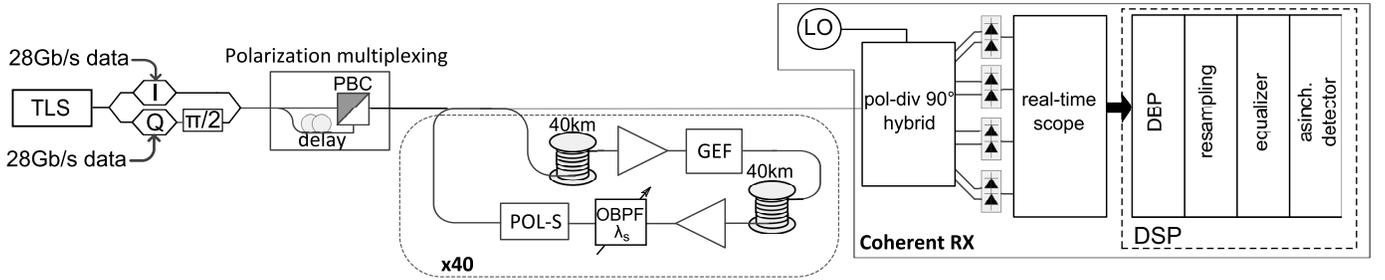}\caption{Experimental setup. \label{fig3}}
\end{figure*}

Finally, also the latency due to the different algorithms is of fundamental
importance for their real-time implementation. In this case, as almost
all the filtering operations involved in the nonlinear sub-step (\ref{eq:passo_nonlineare_ESSFM})
can be executed in parallel, we can assume that the latency of the
ESSFM is almost independent of $N_{c}$ and equal to the latency of
the SSFM. Moreover, the latency due to the whole nonlinear sub-step
can be considered negligible compared to that due to the FFT. Therefore,
we can assume that the total latency of the three algorithms depends
on the total number of cascaded FFTs (a pair for the FFE and $N_{s}$
pairs for the SSFM and ESSFM ) and on their size (equal to the block
length $N$, which depends on the link length and is the same for
all the algorithms ). Again, an absolute and accurate estimate of
the latency is beyond the scope of this work. However, we can compare
the various algorithms by assuming that the latency induced by \emph{each
step} of the SSFM or ESSFM equals that induced by the FFE.

In conclusion, the required number of steps $N_{s}$ can be taken
as a meaningful figure of merit to measure the complexity, latency,
and power consumption of the SSFM and ESSFM algorithms: taking a simple
FFE for bulk dispersion compensation (in dispersion-unmanaged links)
as a basis for comparison, the latency of the SSFM and ESSFM is about
$N_{s}$ times that of the FFE, while their complexity and power consumption
are slightly more than $N_{s}$ times that of the FFE, as indicated
in Fig.\,\ref{fig2}. Therefore, a saving in the number of steps
$N_{s}$ provided by the ESSFM compared to the SSFM (as experimentally
demonstrated in the next section) translates into a proportional saving
in terms of latency, and an almost proportional saving in terms of
complexity and power consumption.

\vspace*{1.5ex}

\section{Experimental results}

The experimental setup employed to compare the performance of the
SSFM and ESSFM algorithms is depicted in Fig.\,\ref{fig3}. An optical
carrier, generated by a 100\,kHz-linewidth tunable laser source (TLS),
is modulated by means of an integrated double nested Mach Zehnder
modulator (IQ-MZM). Two PRBS signals of length $2^{11}-1$ at $\unit[28]{Gb/s}$
are applied to the in-phase (I) and quadrature (Q) port of the modulator
to obtain a 56\,Gb/s QPSK optical signal. Polarization multiplexing
is finally emulated through a 50/50 beam splitter, an optical delay,
and a polarization beam combiner (PBC), obtaining a 112\,Gb/s PM-QPSK
optical signal. A recirculating loop is used to emulate transmission
over long distances. The loop is composed by two spans of 40\,km
of standard single-mode fiber, each one followed by an erbium-doped
fiber amplifier (EDFA). A gain equalization filter (GEF) is used to
equalize the distortions due to the amplifier gain profile and a polarization
scrambler (POL-S) is included in the loop to emulate random polarization
rotations along the link. At the receiver, the optical signal is detected
by employing coherent phase- and polarization-diversity detection
and setting the local oscillator (LO) at the same nominal wavelength
of the transmitter TLS (with $\pm\unit[2]{GHz}$ accuracy). The received
optical signal is mixed with the LO through a polarization-diversity
90\textdegree{} hybrid optical coupler, whose outputs are sent to
four couples of balanced photodiodes. The four photodetected signals
are sampled and digitized through a 20\,GHz 50\,GSa/s real-time
oscilloscope in separate blocks of one million samples at a time.
Each block of samples is processed off-line according to the scheme
of Fig.\,\ref{fig3}. Bulk dispersion compensation (with a frequency-domain
FFE) or DBP based on the SSFM or ESSFM algorithm is performed on signal
samples taken at the original sampling rate (about 1.8 sample per
symbol). Then, after digital resampling at two samples per symbol,
a butterfly equalizer is employed to adaptively compensate for polarization
mode dispersion and residual chromatic dispersion. Finally, asynchronous
detection is employed (at symbol rate) to account for phase noise
and a possible frequency offset and to make decisions as in \cite{cugini2013push}.
The first 100000 received samples are used to optimize the ESSFM coefficients,
while bit-error rate (BER) is measured on the remaining samples. The
ESSFM coefficients are optimized by using the output of the SSFM algorithm
with multiple steps/span as a target and minimizing the mean square
error (MSE) with respect to it.%
\footnote{This approach can be employed even in a real system, as the optimization
can be done off-line when designing the link. A more practical (and
possibly accurate) approach is that of minimizing the MSE between
the output samples (after DBP, equalization, and phase-noise/frequency-offset
compensation) and the transmitted symbols, as suggested in \cite{Secondini:ECOC14}.
This, however, needs some care to handle possible interactions with
the convergence of the butterfly equalizer and is left to a future
investigation.%
} BER values are finally obtained by averaging over 5 different blocks
of samples.

\begin{figure}
\noindent \centering{}\includegraphics[width=1\columnwidth]{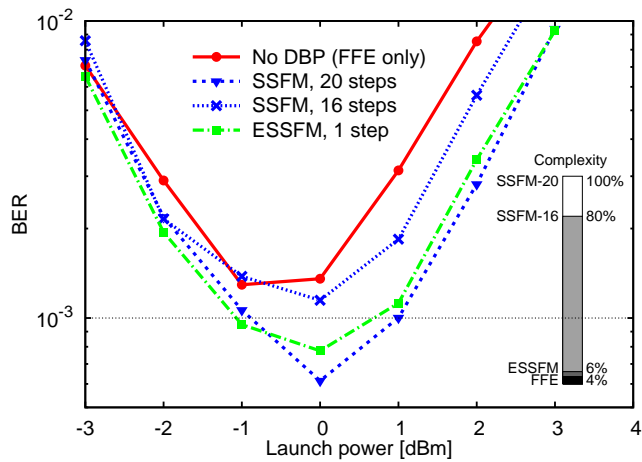}\caption{BER versus launch power for different DBP implementations at a distance
of 3200\,km. Inset: relative computational complexity of the various
algorithms, with the 20-step SSFM equal to 100\%.\label{fig4}}
\end{figure}

The performance and complexity of the SSFM and ESSFM algorithm are
compared at a transmission distance of 3200\,km, at which the system
operates with a BER above an arbitrary prescribed threshold of $10^{-3}$
without DBP (with FFE only). The BER versus launch power obtained
without DBP (replaced by the FFE for dispersion compensation), with
the SSFM, and with the ESSFM algorithm is shown in Fig.\,\ref{fig4}.
At this distance, a channel memory $M\simeq1024$ samples and a nearly
optimal FFT size $N=8192$ are taken. Different number of steps $N_{s}$
for the SSFM and ESSFM algorithms are considered. For the ESSFM algorithm,
$N_{c}$ is selected to provide a good trade-off between performance
and complexity. For the system without DBP, the minimum BER is obtained
at a launch power of -1\,dBm and is higher than the prescribed threshold.
When including DBP based on the standard SSFM algorithm, at least
20 steps (one step each four spans) are required to obtain $\mathrm{BER}<10^{-3}$
(at a launch power of 0\,dBm). On the other hand, when the ESSFM
algorithm is used to implement DBP, the prescribed BER can be achieved
(already at -1\,dBm of launch power) with just a single step for
the whole link and $N_{c}=32$ coefficients. The total number of real
multiplications per received sample is 128 for the FFE, 179 for the
ESSFM, 2286 for the 16-step SSFM, and 2857 for the 20-step SSFM. The
relative complexity (and power consumption) of the various algorithms
is shown in the inset of Fig.\,\ref{fig4}, taking the 20-step SSFM
as a reference. By employing the ESSFM, the overall complexity and
power consumption are reduced by a factor of 16 with respect to a
conventional SSFM with the same performance, and latency by a factor
of 20.\vspace*{0.5ex}

\section{Conclusions}

Low-complexity DBP based on the ESSFM algorithm has been experimentally
demonstrated by backpropagating a 112\,Gb/s PM-QPSK signal through
a 3200\,km dispersion-unmanaged link. A target BER of $10^{-3}$
has been achieved with a single DBP step, with a 16 times lower complexity
and 20 times lower latency than conventional DBP. This means that
ESSFM allows for complexity, latency, and power-consumption comparable
with those required by standard feedforward equalization for chromatic
dispersion compensation.\vspace*{0.5ex}

\section*{Acknowledgment}

This work was supported in part by the Italian MIUR under the FIRB
project COTONE and by the EU FP-7 G\'{E}ANT project COFFEE.

\end{document}